\documentclass[reprint,superscriptaddress,amsmath,amssymb,aps,prx,longbibliography]{revtex4-2}

\usepackage[utf8]{inputenc}
\usepackage{amsfonts}
\usepackage{xcolor}
\usepackage{longtable}
\usepackage[english]{babel}

\usepackage[pdfstartview=FitH,
            breaklinks=true,
            bookmarksopen=false,
            bookmarksnumbered=true,
            colorlinks=true,
            linkcolor=black,
            citecolor=black,
            urlcolor=black,
            pdftitle={mlTCRdist},
            pdfauthor={Andrew Pyo et al.},
            ]{hyperref}

\usepackage{mathtools}
\DeclarePairedDelimiter{\expval}{\langle}{\rangle}

\begin{document}

\title{
Data-driven Discovery of Biophysical T Cell Receptor Co-specificity Rules
}

\author{Andrew G. T. Pyo}
\affiliation{Department of Physics, Princeton University, Princeton, NJ, USA}
\author{Yuta Nagano}
\affiliation{Division of Infection and Immunity, University College London, London, UK}
\affiliation{Division of Medicine, University College London, London, UK}
\author{Martina Milighetti}
\affiliation{Division of Infection and Immunity, University College London, London, UK}
\affiliation{Cancer Institute, University College London, London, UK}
\author{James Henderson}
\affiliation{Division of Infection and Immunity, University College London, London, UK}
\affiliation{Institute for the Physics of Living Systems, University College London, London, UK}
\author{Curtis G. Callan Jr.}
\affiliation{Department of Physics, Princeton University, Princeton, NJ, USA}
\affiliation{Département de Physique, École Normale Supérieure, Paris, France}
\author{Benny Chain}
\affiliation{Division of Infection and Immunity, University College London, London, UK}
\affiliation{Department of Computer Science, University College London, London, UK}
\author{Ned S. Wingreen}
\affiliation{Lewis-Sigler Institute for Integrative Genomics, Princeton University, Princeton, NJ, USA}
\affiliation{Department of Molecular Biology, Princeton University, Princeton, NJ, USA}
\author{Andreas Tiffeau-Mayer}
\affiliation{Division of Infection and Immunity, University College London, London, UK}
\affiliation{Institute for the Physics of Living Systems, University College London, London, UK}
\date{\today}

\begin{abstract}

The biophysical interactions between the T cell receptor (TCR) and its ligands determine the specificity of the cellular immune response. However, the immense diversity of receptors and ligands has made it challenging to discover generalizable rules across the distinct binding affinity landscapes created by different ligands. Here, we present an optimization framework for discovering biophysical rules that predict whether TCRs share specificity to a ligand. Applying this framework to TCRs associated with a collection of SARS-CoV-2 peptides we systematically characterize how co-specificity depends on the type and position of amino-acid differences between receptors. We also demonstrate that the inferred rules generalize to ligands highly dissimilar to any seen during training. Our analysis reveals that matching of steric properties between substituted amino acids is more important for receptor co-specificity than the hydrophobic properties that prominently determine evolutionary substitutability. Our analysis also quantifies the substantial importance of positions not in direct contact with the peptide for specificity. These findings highlight the potential for data-driven approaches to uncover the molecular mechanisms underpinning the specificity of adaptive immune responses.

\end{abstract}

\maketitle

Cellular immunity relies on the specific recognition of target molecules by T cells, mediated by the binding of the T cell receptor (TCR) to specific peptide-major histocompatibility complexes (pMHCs) \cite{Davis1988CellReceptor,chi2024principles}. Given the pivotal role of T cells in the adaptive immune system, an ability to predict the specificity of a TCR {\it in silico} from its sequence would have many applications in disease diagnosis, surveillance, and treatment. 
Recent experimental advances \cite{dash_quantifiable_2017, glanville_identifying_2017,dobson2022antigen,joglekar2019t,sureshchandra2024tissue} and the creation of databases cataloging known TCR-pMHC pairings \cite{bagaev_vdjdb_2020, Tickotsky2017, Vita2018, Nolan2020} have thus propelled significant research efforts aimed at predicting TCR specificity using machine learning \cite{gielis_detection_2019, lin_rapid_2021, fischer_predicting_2020, jokinen_predicting_2021,  croce2024deep, Wang2024, meynard2024tulip, zhang2024epitope, nagano2025contrastive}.

The premise behind using machine learning to predict pairings between the hypervariable receptors and their ligands is that there are learnable, generalizable biophysical rules that govern these interactions \cite{dash_quantifiable_2017,glanville_identifying_2017}. 
While machine learning models are able to accurately identify alternative binders to pMHCs for which good quality high-volume TCR data exists, an ability to generalize to pMHCs unseen during training has yet to be demonstrated in independent benchmarks \cite{nielsenlessons2024}.

In high-dimensional multi-class classification problems a potentially easier path towards generalizable learning begins with learning similarity relationships between class members \cite{bellet2013survey,vinyals2016matching}. Once a suitable metric or representation tailored to the general nature of the classification problem is learned, it can then be adapted with little training data to new classes. However, to date the most popular sequence similarity metrics for TCRs are heuristic and their parameters have not been learned from TCR sequence data. For instance, TCRdist  \cite{dash_quantifiable_2017} scores amino-acid similarity based on the BLOcks SUbstitution Matrix (BLOSUM). This matrix is based on substitution frequencies observed in general protein evolution \cite{henikoff_amino_1992}, and it is unclear how well these describe the impact of amino-acid changes on TCR-pMHC binding. Additionally, existing structures suggest that factors beyond the identity of substituted amino acids, such as the position of the substitution, likely influence the impact of sequence changes \cite{milighetti2021predicting}.

\begin{figure*}
\centering
\includegraphics[scale=0.4]{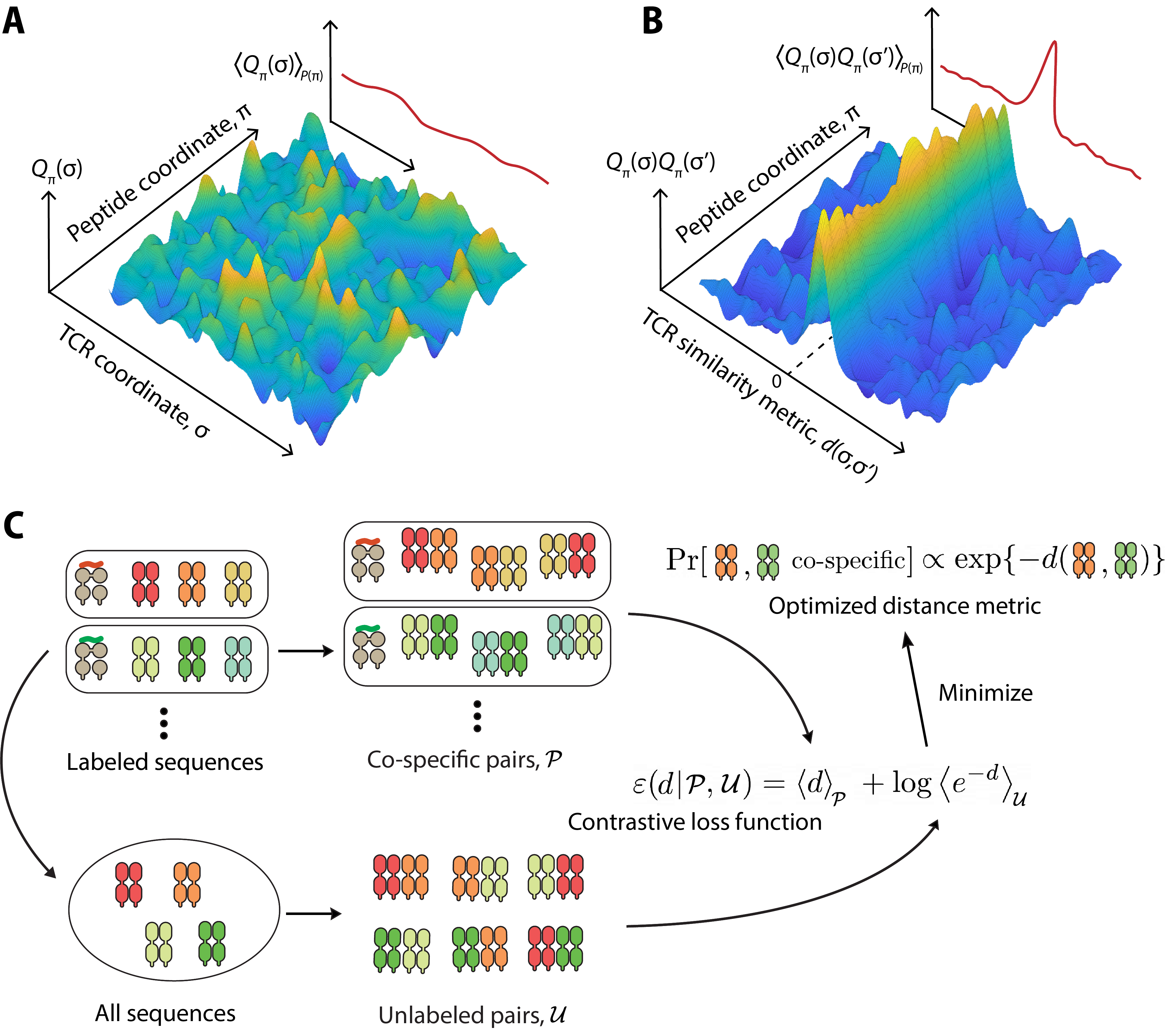}
\caption{{\bf Contrastive learning of rules that generalize across complex receptor-ligand maps.} \textbf{A)} Cartoon disordered ``landscape" of selection factors $Q_\pi(\sigma)$, which describe the varying binding affinities of TCRs $\sigma$ to ligands $\pi$. Averaging selection factors over ligands, $\expval{Q_\pi(\sigma)}_{P(\pi)}$, will lead to a largely flat marginal distribution (inset), unless we are able to restrict the average to ligands with highly similar TCR landscapes. \textbf{B)} Cartoon ``landscape" of co-selection factors $Q_\pi(\sigma)Q_\pi(\sigma')$, which relate to the probability with which two TCRs are specific to the same ligand, as a function of TCR similarity $d(\sigma, \sigma')$. In contrast to the previous scenario, co-selection factors have a non-trivial marginal distribution when averaged across unrelated ligands, $\expval{Q_\pi(\sigma)Q_\pi(\sigma')}_{P(\pi)}$ (inset). \textbf{C)} Sketch of the supervised contrastive learning paradigm. Experimentally-determined pairs of TCR sequences binding the same ligand make up the set of co-specific pairs $\mathcal{P}$, and all pairs of TCR sequences irrespective of specificity make up the set of unlabeled pairs $\mathcal{U}$. We optimize parameters $\theta$ of a family of distance metrics $d_\theta$ which minimize a contrastive loss (Eq. \ref{eq:loss}). The loss function minimizes the average distance between pairs in $\mathcal{P}$, while pushing apart unlabeled pairs in $\mathcal{U}$. Optimized parameters $\theta^\star$ relate the sequence similarity between a pair of TCRs to the probability with which they share ligand specificity and provide biophysical insights into feature importance.} 
\label{figsketch}
\end{figure*}

To address these challenges, we propose a pseudo-likelihood maximization approach to learn the biophysical rules that govern TCR co-specificity directly from data. Concretely, our aim is to learn a distance metric $d(\sigma,\sigma')$ that quantifies the probability that two TCRs, $\sigma$ and $\sigma'$, bind to the same ligand (pMHC). To accomplish this, we expand on our ongoing efforts to quantify the statistics of TCR sequence similarity from available TCR-pMHC specificity data \cite{mayer_measures_2023,tiffeau2024unbiased,henderson_limits_2024} by replacing predefined notions of sequence similarity with learnable metrics. To learn this metric we apply contrastive learning -- a framework that we and others have previously used to train deep learning models for predicting TCR specificity \cite{drost_metcrs_2022,pertseva_tcr_2024,nagano2025contrastive,zhang2024epitope} -- to optimize sequence alignment weights to predict TCR co-specificity. We reason that the simplicity of these metrics might favor the inference of generalizable rules from limited data, which has remained a challenge in more flexible deep learning models using this supervised contrastive approach  \cite{nagano2025contrastive}.

Our results show that co-specificity rules inferred from a large-scale dataset on SARS-CoV-2 specific TCRs \cite{Nolan2020} generalize with striking breadth. These rules accurately predict TCR co-specificity even for pMHCs that are highly dissimilar (differing by $\geq$ 6 amino acids) from the nearest peptide in the training set, demonstrating an extrapolation ability that has so far remained out of reach for sequence-based machine learning approaches \cite{nielsenlessons2024, meynard2024tulip}.
Beyond the empirical findings, we provide a theoretical perspective on how contrastive learning is related to pseudolikelihood methods and intuition for why it might identify reproducible order within disordered receptor-ligand landscapes, where traditional inference methods often falter. Altogether, our findings position contrastive learning as a powerful and broadly applicable strategy for developing statistical theories of complex receptor-ligand maps from limited data.

\section{Two-point statistics identify order in receptor-ligand maps}

To motivate our framework, it is useful to review the statistical inference problem posed by TCR-pMHC binding prediction. The ultimate objective is to infer the probability distribution $P_\pi(\sigma)$ of observing TCR sequence $\sigma$ when sampling from the pool of T cells specific to pMHC $\pi$. This distribution can be fitted from known experimentally sampled TCR binders to the pMHC. However, there are only few pMHCs with substantial data, making the inference of such pMHC-specific models challenging.  To address this, it would be appealing to use generalizable rules as priors when inferring the selection landscape for a given pMHC.

One approach involves using the pMHC-specific selection factors $Q_\pi(\sigma)$, such that $P_\pi(\sigma) = Q_\pi(\sigma) P(\sigma)$, where $P(\sigma)$ represents the highly highly non-uniform \cite{murugan_statistical_2012}, but largely universal \cite{sethna2020population} baseline distribution resulting from VDJ recombination \cite{Elhanati2014QuantifyingSelection,mayer_measures_2023,bravi2023transfer}. 
A naive strategy might average selection factors across pMHCs, $\expval{Q_\pi(\sigma)}_{P(\pi)}$ as an initial guess, where $\expval{f(x)}_{P(x)}$ denotes the average of $f(x)$ over the distribution $P(x)$.
Yet, such a ``mean-field" approach would obscure much of the information about TCR-pMHC specificity. 

The binding energy landscapes associated with specific ligands peak in different regions of receptor sequence space (Fig.~\ref{figsketch}A), and averaging over these disordered landscapes is expected to result in a flat and non-informative marginal distribution (Fig.~1A, inset). Indeed, embeddings from masked language models trained to reproduce overall repertoire statistics demonstrate limited transfer learning capability for predicting TCR-pMHC specificity \cite{nagano2025contrastive}. Furthermore, selection factors on observed naive or memory TCR repertoires relative to null expectations are largely driven by pMHC-independent constraints on receptor function, such as those required for proper folding \cite{Elhanati2014QuantifyingSelection}.

To overcome this problem, we have followed a different approach towards identifying reproducible statistical order across receptor-ligand maps, which rests on analyzing two-point statistics of selection $Q_\pi(\sigma)Q_\pi(\sigma')$ \cite{mayer_measures_2023,nagano2025contrastive,henderson_limits_2024, tiffeau2024unbiased}. The two-point statistics describe the likelihood that receptors $\sigma$ and $\sigma'$ are both specific to the same peptide $\pi$ (Fig.~1B). These co-selection factors can have non-trivial averages across unrelated ligands (Fig.~\ref{figsketch}B, inset), even where one-point statistics do not. Intuitively, peak position vary across the selection landscapes, but the size and shape of peaks might not.  Two-point statistics have also been successfully applied in the statistical physics of spin glasses to reveal otherwise hidden statistical order  \cite{edwards1975theory}.

Using this approach, we previously demonstrated that co-selection factors decay predictably with the Levenshtein distance between TCR sequences, independent of the specific ligand \cite{mayer_measures_2023}. These findings suggest that generalizable statistical order exists across receptor-ligand maps, which can inform metrics for TCR sequence analysis.

\section{Supervised contrastive learning as pseudo-likelihood maximization} 

\subsection{Derivation of a loss function}

We propose fitting available co-specificity data by optimizing parameters $\theta$ of a family of TCR sequence similarity metrics $d_\theta(\sigma, \sigma')$. The fitting is based on a supervised contrastive loss function, similar to others used in the machine learning literature \cite{khosla_supervised_2021}. In the following we provide a derivation of this loss function that clarifies its connection to pseudo-likelihood maximization -- a technique used by statistical physicists to infer Potts models for protein families \cite{ekeberg2013improved}.

The experimental data takes the form of a collection of TCRs $\{ \sigma_1, \sigma_2, ..., \sigma_N \}$, where each sequence $\sigma_i$ is associated with a cognate pMHC $\pi_i$ (Fig.~\ref{figsketch}C). To find an optimal metric we propose maximizing the pseudo-likelihood of all observed pairs of co-specific TCRs:
\begin{equation}
\label{eq:pseudo}
    \mathcal{L} = \prod_{\substack{i\ne j \\ \pi_i = \pi_j}}Q_{\pi_i}(\sigma_i)Q_{\pi_i}(\sigma_j)P(\sigma_i)P(\sigma_j).
\end{equation}
We have shown previously that co-selection factors between randomly chosen pairs of sequences at a given Levenshtein distance $d_\mathrm{LD}(\sigma, \sigma') = \Delta$,
\begin{equation}
\expval{Q_\pi(\sigma)Q_\pi(\sigma') }_{P(\sigma, \sigma' | d_\mathrm{LD}(\sigma, \sigma')=\Delta)},
\end{equation}
decay exponentially at small distances $\Delta$ with a typical, ligand-independent length scale \cite{mayer_measures_2023}. 
In fitting a metric to maximize the pseudo-likelihood defined in Eq.~\ref{eq:pseudo} we thus focus on capturing the average behavior of co-selection factors across pMHCs,
\begin{equation}
    Q_{\pi_i}(\sigma)Q_{\pi_i}(\sigma') \approx\,
    Q(\sigma,\sigma') \equiv \expval{Q_\pi(\sigma)Q_\pi(\sigma')}_{P(\pi)},
\end{equation}
according to an exponential ansatz,
\begin{equation} \label{eq:qpropto}
Q(\sigma, \sigma') \propto e^{-d_\theta(\sigma,\sigma')}.
\end{equation}
 Although outside the scope of our study, it is clear that this metric could be improved for prediction on specific targets by exploiting ligand-dependent variation in these rules \cite{tiffeau2024unbiased,henderson_limits_2024}, as we have previously demonstrated for protein language models \cite{nagano2025contrastive}. Here, inference is instead by design restricted to co-specificity rules that generalize across ligands.

To complete the specification of the optimization problem, it remains to determine the proportionality constant in Eq.~\ref{eq:qpropto}. We use a normalization condition $\expval{Q(\sigma,\sigma')}_{P(\sigma)P(\sigma')} = 1$, such that
\begin{equation}
\label{eq:Q}
    Q(\sigma,\sigma') = \frac{e^{-d_\theta(\sigma,\sigma')}}{\expval{e^{-d_\theta(\sigma,\sigma')}}_{P(\sigma)P(\sigma')}}.
\end{equation}
With sufficient data, the average in the denominator can be approximated by summation over all pairs of TCRs in the training data. We call this set of pairs the unlabeled set, $\mathcal{U} = \{(\sigma_i,\sigma_j)|i\ne j,\forall i,j\}$, as it ignores the associated pMHC labels. This means approximating the partition function with respect to $P(\sigma)P(\sigma')$ by an average with respect to the empirical distribution of unlabeled pairs $\mathcal{U}(\sigma,\sigma') = |\mathcal{U}|^{-1}\sum_{(s,s')\in\mathcal{U}}\delta_{\sigma,s}\delta_{\sigma',s'}$. Here, $|\mathcal{U}| = N(N-1)/2$ denotes the cardinality of the unlabeled set, and $\delta_{x,y}$ the Kronecker delta, $\delta_{x,y} = 1$ for $x=y$ and $\delta_{x,y} = 0$ otherwise. 
We note that an alternative approach would be to calculate the average only with respect to pairs of TCRs with differing pMHC specificity or by using a large set of independently acquired TCRs, say from an unsorted blood sample. However, in line with prior work \cite{wang_understanding_2022, nagano2025contrastive} we found that the inclusion of co-specific pairs in the evaluation of the denominator has a regularizing effect on the inference procedure.

To simplify notations, we similarly define the set of positive pairs, $\mathcal{P} = \{(\sigma_i,\sigma_j)|\pi_i = \pi_j,\forall i,j\}$, and the empirical probability of observing TCR pair $(\sigma,\sigma')$ in the set of co-specific pairs, $\mathcal{P}(\sigma,\sigma') = |\mathcal{P}|^{-1}\sum_{(s,s')\in\mathcal{P}}\delta_{\sigma,s}\delta_{\sigma',s'}$.
Plugging all definitions into Eq. \ref{eq:pseudo} allows us to write the negative log-pseudo-likelihood per positive pair as
\begin{equation}
    -\frac{\log \mathcal{L}}{|\mathcal{P}|} = \langle d_\theta(\sigma,\sigma) \rangle_{\mathcal{P}(\sigma, \sigma')}+\log \expval{e^{-d_\theta(\sigma,\sigma')}}_{\mathcal{U}(\sigma, \sigma')} + c,
\end{equation}
where $c = - \langle \log P(\sigma)P(\sigma') \rangle_{\mathcal{P}(\sigma,\sigma')}$ is a term that is independent of the distance metric $d_\theta$. 
Optimal parameters $\theta^\star$ maximize the pseudo-likelihood (Eq. \ref{eq:pseudo}), or equivalently, minimize terms in the negative log-likelihood dependent on $d_\theta$, which yields the following loss function
\begin{equation}
\label{eq:loss}
    \varepsilon(\theta|\mathcal{P},\mathcal{U}) = \langle d_\theta(\sigma,\sigma') \rangle_{\mathcal{P}(\sigma,\sigma')}+\log\langle e^{-d_\theta(\sigma,\sigma')} \rangle_{\mathcal{U}(\sigma,\sigma')}.
\end{equation}
Fig.~\ref{figsketch}C summarizes in a flowchart how the sets $\mathcal{P}$ and $\mathcal{U}$ of co-specific and unlabeled sequence pairs are constructed.  

\subsection{Analytical solution for a simple case}

To build intuition, we minimize Eq.~\ref{eq:loss} in closed form for a simple distance metric. We consider distance metrics defined in terms of a discrete-valued function $g$ that maps each pair of TCRs to a finite set of $K$ integers (e.g. for $K=2$ a simple case would be: $g(\sigma,\sigma')=1$ if there is an insertion or deletion and $g(\sigma,\sigma')=0$ for an amino-acid substitution). Given the integer-valued function $g$, a family of distance metrics with trainable parameters can be defined by
\begin{equation}
\label{eq:analytic_d}
    d_{g}(\sigma,\sigma') = \sum_{i=1}^K \theta_i \delta_{g(\sigma,\sigma'),i},
\end{equation}
where $\theta_1, \theta_2,..., \theta_K$ are parameters assigned to the $K$ possible value of $g$, and $\delta_{a,b} = 1$ for $a=b$ and $\delta_{a,b}=0$ otherwise is the Kronecker delta function. Substituting Eq.~\ref{eq:analytic_d} into the loss function (Eq. \ref{eq:loss}) yields
\begin{equation} \label{eqlossg}
    \varepsilon(\theta|\mathcal{P},\mathcal{U}) = \sum_{i=1}^K \theta_i \expval{\delta_{g(\sigma,\sigma'),i}}_{\mathcal{P}(\sigma,\sigma')} + \log Z,
\end{equation}
where $Z = \left\langle e^{-\sum_i \theta_i\delta_{g(\sigma,\sigma'),i}}\right\rangle_{\mathcal{U}(\sigma,\sigma')}$ is a normalizing constant.
At the optimal $\theta_i^\star$, we have $\partial\varepsilon/\partial \theta_i = 0$. Taking the partial derivative of Eq.~\ref{eqlossg} with respect to $\theta_i$ yields
\begin{equation} \label{eqpartial}
\frac{\partial \epsilon}{\partial \theta_i} = \expval{\delta_{g(\sigma,\sigma'),i}}_{\mathcal{P}(\sigma,\sigma')}
- \frac{e^{-\theta_i}}{Z}\expval{\delta_{g(\sigma,\sigma'),i}}_{\mathcal{U}(\sigma,\sigma')}.
\end{equation}
Solving for $\theta_i^\star$ by setting Eq.~\ref{eqpartial} to zero we derive 
\begin{equation}
    \theta_i^\star = -\log \frac{\expval{\delta_{g(\sigma,\sigma'),i}}_{\mathcal{P}(\sigma,\sigma')}}{\expval{\delta_{g(\sigma,\sigma'),i}}_{\mathcal{U}(\sigma,\sigma')}} - \log Z.
\end{equation}
This result shows that minimizing the contrastive loss function (Eq.~\ref{eq:loss}) sets parameters based on the log-ratio between observed occurrences of particular features among positive pairs and a null expectation calculated across unlabeled pairs.

This analytical solution shows that contrastive learning using Eq.~\ref{eq:loss} is closely connected to how BLOSUM and related evolutionary amino-acid substitution matrices  are traditionally defined in the bioinformatics literature \cite{altschul1991amino,henikoff_amino_1992}. These matrices are also constructed by calculating log-odds ratios between observed and expected frequencies of amino acid pairs. However, contrastive learning extends this by accommodating more complex distance metrics that incorporate multiple substitution properties. Furthermore, empirical frequencies across unlabeled TCR pairs are used in the denominator, rather than assuming factorization as in BLOSUM. This is a more appropriate choice for TCRs, where VDJ recombination introduces correlations between sites even in the naive repertoire \cite{murugan_statistical_2012}.

\section{Data preparation}
\label{sec_data_preparation}

In order to learn rules that generalize to unseen pMHCs, we curated the training data to eliminate confounding factors and to reduce the complexity of the problem. We decided to restrict training to a single dataset of TCRs with experimentally annotated pMHC specificity to exclude confounding by data origin. Consequently, we used a large-scale dataset on TCRs with specificity to SARS-CoV-2 peptides obtained using a multiplexed assay called MIRA \cite{Nolan2020}. From this data we only used pairs of TCR CDR3$\beta$ sequences that differ by a single edit step to maximize the signal-to-noise ratio given potentially misannotated sequences \cite{mayer_measures_2023}. Moreover, in both training and testing, we only considered pairs of sequences with identical V genes, to ensure that learned rules only reflected amino-acid substitutions in the CDR3$\beta$ chain. We also excluded the (very small) number of pairs involving substitutions of cysteine to reduce overfitting, due to the rarity of cysteines within the CDR3 junction following negative thymic selection \cite{Elhanati2014QuantifyingSelection}.

To reduce overfitting to less commonly observed sequence lengths, we restricted our analysis to TCR CDR3$\beta$ sequences with lengths between 11 and 15. These represented 72\% of the sequences in the MIRA database and 76\% of sequences in the VDJdb database. From the MIRA database, we identified 26 sets of TCRs specific to unique pMHCs, each containing at least 1000 TCR sequences. To mitigate the impact of varying sizes of specificity groups, we equalized all group sizes by subsampling without replacements to a 1000 sequences per pMHC. To still learn on all available data, we repeated the procedure 15 times and averaged performance across these training batches. Among the 26,000 TCRs included per training batch, we identified on average $\approx 6000$ pairs of single-edit TCRs.

We evaluated the inferred metric using TCRs specific to hold-out pMHCs from the MIRA database. To this end, we collated sets of co-specific sequence pairs from 24 specificity groups, each containing between 400 and 999 TCR sequences, that were excluded during training. We again used subsampling to address class imbalance, generating 15 sets of 400 sequences for each specificity group. Furthermore, to assess generalization to TCR pairs differing by multiple edits, we identified TCR pairs with edit distances ranging from 1 to 6 in the same data set.

Using these data, we chose to test the ability of metrics to discriminate between co-specific (i.e.~$\in \mathcal{P}$) and cross-specific TCR pairs (i.e.~$\in \mathcal{C} = \{(\sigma_i,\sigma_j)|\pi_i\ne \pi_j, \ \forall i,j\})$. We note that this task represents an instance of positive-unlabeled classification \cite{bekker2020learning}, as cross-specific pairs are not tested experimentally for co-specificity. We have shown previously that about one in ten single edits maintain co-specificity \cite{mayer_measures_2023}, setting an upper bound on achievable classification accuracy. Nonetheless, we chose this task for its simplicity. Additionally, when considering TCR pairs differing by multiple edits the prevalence of co-specific pairs decreases, reducing the impact of this limitation.

To externally validate the predictive power of the inferred rules, we used data from VDJdb \cite{Goncharov2022}, excluding any pMHCs present in the MIRA dataset. In addition to TCR$\beta$ sequences, VDJdb includes TCR$\alpha$ sequences for certain pMHCs, which we used to further test the biological generalizability of the inferred co-specificity rules. For the $\beta$ chain validation task, we selected TCRs from 131 specificity groups, each containing at least 15 unique sequences. As before, this threshold was chosen to maximize the number of co-specific pairs included in the analysis after balancing the data across groups. For the $\alpha$ chain validation task, we identified 49 specificity groups meeting the same threshold of 15 sequences. Consistent with our previous methodology, we employed subsampling and identified TCR pairs with edit distances ranging from 1 to 6 across 15 balanced testing batches.

\begin{figure*}
\centering
    \includegraphics[scale=0.5]{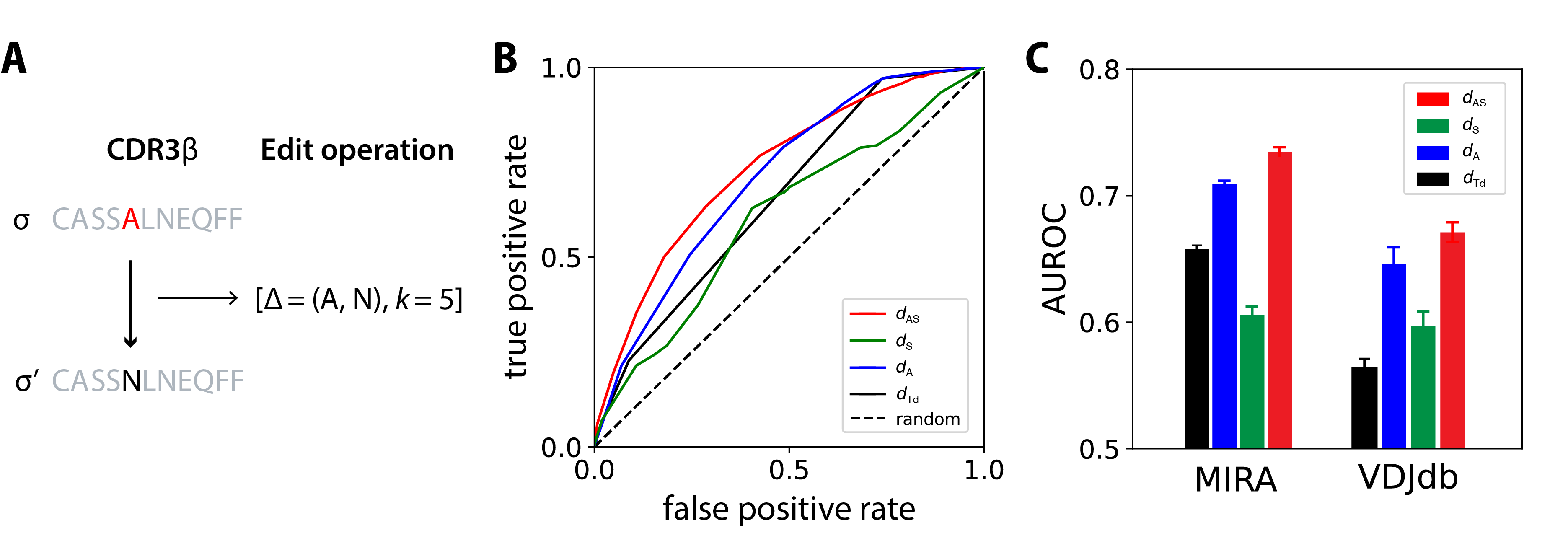}
\caption{ {\bf Learning of a CDR3$\beta$ distance metric that generalizes to unseen ligands.} \textbf{A)} Distance metrics are defined in terms of the edit steps between two CDR3$\beta$ sequences. Each edit step includes the length of sequence $\ell$, position of substitution $\kappa$, and the identity of substitution $\Delta$. \textbf{B)} Receiver Operating Characterics (ROC) curves for identifying co-specific relative to cross-specific (i.e., $\in \mathcal{C} = \{(\sigma_i,\sigma_j)|\pi_i\ne \pi_j, \ \forall i,j\}$) TCR CDR3$\beta$ pairs from the MIRA test set that differ by a single edit using different TCR similarity metrics. The metrics include TCRdist ($d_\mathrm{Td}$, black) as a baseline, as well as a series of metrics optimized using the mathematical framework proposed in this study. These metrics involve substitution type (which amino acids are substituted) and/or substitution site (which position along the CDR3 sequence is substituted). Specifically, the curves show metrics using only type ($d_\mathrm{A}$, blue), only site ($d_\mathrm{S}$, green), or using both jointly ($d_\mathrm{AS}$, red). Curves are averaged over 15 subsampled test sets, each equally balanced across pMHC groups. \textbf{C)} Areas under the ROC (AUROC) for the MIRA test set and an external validation set (VDJdb). Error bars indicate the standard error over resampled testing batches. Note that due cross-reactivity some cross-specific pairs also exhibit co-specificity, which bounds achievable AUROCs below one.} 
\end{figure*}

\section{Inference of optimal sequence similarity metrics}

\subsection{Overview, notations, and baseline metric}

To provide proof-of-concept for the applicability of our framework to this dataset, we applied our inference procedure to optimize parameters within a family of simple alignment-based distance metrics.  Alignment-based metrics of TCR sequence similarity have been widely used to predict TCR co-specificity \cite{dash_quantifiable_2017, Chronister2021, glanville_identifying_2017,mayer-blackwell2021tcr}. Among these, one of the most widely used metrics is TCRdist \cite{dash_quantifiable_2017}, which calculates distances between TCRs based on BLOSUM62-derived amino-acid dissimilarity penalties. TCRdist furthermore optionally trims CDR3 sequences to exclude residues at the edges from the mismatch calculation, as a simple form of non-uniform weighting of mismatches according to position inspired by structural evidence showing that these flanking residues rarely make direct contact with the peptide \cite{dash_quantifiable_2017,mayer-blackwell2021tcr}. Building on the principles of TCRdist, we focused on the inference of parameters relating to two critical physical factors influencing the effects of changes in the TCR hypervariable region: the identity of the substituted amino acids and the position of the substitution.

Specifically, we define distance metrics in terms of edit steps, which characterize the difference between a pair of TCR sequences $\sigma$ and $\sigma'$ by a series of substitutions, deletions, and insertions (Fig.~2A). The simplest among these metrics, the total unweighted number of edit steps required to transform $\sigma$ to $\sigma'$, is known as the Levenshtein distance \cite{Lev_66}. In the following we describe the $i$th edit step between $\sigma$ and $\sigma'$ by $\Delta_i(\sigma,\sigma')$, the identity of substituted amino acids, and by $k_i(\sigma,\sigma')$, the substitution position along the TCR sequence (Fig.~2A). 

In this notation, the baseline metric, TCRdist without trimming, is given by $d_\mathrm{Td}(\sigma,\sigma') = \sum_i \tilde{d}_{\mathrm{Td}}(\Delta_i)$, where $\tilde{d}_{\mathrm{Td}}(\Delta_i) = \min(4, 4 - B_{62}(\Delta_i))$ if the $i^\mathrm{th}$ edit step is an amino-acid substitution, and $\tilde{d}_{\mathrm{Td}}(\Delta_i) = 8$ if the $i^\mathrm{th}$ edit step is an insertion/deletion (see Fig.~S1A). Here, $B_{62}$ denotes the BLOSUM62 matrix. In the remainder of this section, we use $d_{Td}$ as a baseline for comparisons.

\subsection{Learning an amino-acid substitution matrix}

To test whether an amino-acid substitution matrix directly learned from TCR data would outperform the BLOSUM-derived TCRdist score, we used our framework to train the following simple metric
\begin{equation}
\label{eq:dA_metric}
    d_\mathrm{A}(\sigma,\sigma') = \sum_i A(\Delta_i),
\end{equation}
where $A$ is a matrix of amino-acid substitution weights (Fig.~S1B). We included a blank (`$-$') in the amino-acid alphabet to treat insertion/deletion as a substitution involving `$-$'.

To evaluate the performance of $d_\mathrm{A}$, we generated pairs of TCR CDR3$\beta$ sequences from the MIRA database that differed by a single edit step and belonged to a specificity group not included in the training data. The performance of distance metrics was quantified using a co-specific pair identification test, in which co-specific TCRs were treated as positive pairs, while TCR pairs across different specificity groups were treated as negative pairs as described before. The receiver operating characteristic (ROC) curves for different models shows that $d_\mathrm{A}$ outperforms $d_\mathrm{Td}$ (Fig.~2B). Furthermore, the learned substitution matrix also better classifies co-specific TCR CDR3$\beta$ sequence pairs from the VDJdb database (Fig.~2C), as assessed by the area under the receiver operating curve (AUROC). Both tasks involve TCRs specific to unseen pMHCs, which in the case of VDJdb additionally originate from pathogens other than SARS-CoV-2 which was used during training.

\subsection{Substitution position influences TCR co-specificity}

We next investigated whether the position of substitution (independently of the identity of the substituted amino acids) determines the likelihood of TCR co-specificity, as suggested by structures \cite{glanville_identifying_2017,milighetti2021predicting}. To address this question, we trained a distance metric that considers only the position of substitution:
\begin{equation}
\label{eq:dS_metric}
    d_\mathrm{S}(\sigma, \sigma') = \sum_i S(k_i, \ell),
\end{equation}
where \( S \) represents a matrix of site weights dependent on \( k_i \), the position of substitution, and \( \ell(\sigma, \sigma') = \max \left(|\sigma|, |\sigma'|\right) \), the aligned length of the TCR pair (Fig.~S2A).

Results on the testing task shows that $d_\mathrm{S}$ has some predictive power on unseen pMHCs (Fig.~2B), and even outperforms $d_\mathrm{Td}$ on the validation task (Fig.~2C). These findings confirm that location matters for the impact of a substitution on TCR specificity.

\subsection{Joint inference of site and identity-dependent substitution weights}

Having established that both the identity of substituted amino acids and the position of substitution influence the probability of co-specificity in a pair of TCRs, we next inferred a model accounting for both factors jointly. Incorporating multiple factors into phenomenological distance metrics is challenging, as it requires ad hoc choices for setting the relative weight given to each factor. In contrast, in our framework, different parameters are jointly optimized during the training process to disentangle their relative influence.

To jointly optimize site and identity-dependent weights while keeping the number of parameters manageable, we defined a  factorized family of distance metrics,
\begin{equation}
\label{eq:dAS_metric}
    d_\mathrm{AS}(\sigma,\sigma') = -\sum_i \log \left[1- \left(1-e^{-\mathcal{S}(k_i,\ell)}\right)\left(1-e^{-\mathcal{A}(\Delta_i)}\right)\right],
\end{equation}
where $\mathcal{S}$ and $\mathcal{A}$ are, respectively, the co-optimized substitution weights that depend on the position of substitution, and the identity of the substitution. At first order, this expression simplifies to the product of weights, $d_{AS}(\sigma,\sigma') \sim \sum_i \mathcal{S}(k_i,\ell) \mathcal{A} (\Delta_i)$, but inspired by prior models of receptor-ligand interactions \cite{kovsmrlj2008thymus,Wang2024,mayer_measures_2023} one can construct a simple toy model to give Eq.~\ref{eq:dAS_metric}  a more biophysical interpretation. We define $P_\mathrm{S}(k, \ell)$ as the probability that a residue at site $k$ on a TCR of length $\ell$ contributes to pMHC specificity, and $P_\mathrm{A}(\Delta)$ as the probability that co-specificity is maintained at a given site following a substitution $\Delta$. Most naively, $P_\mathrm{S}(k, \ell)$ can be interpreted as a contact probability between TCR residue and target peptide, but the importance of a site might also reflect MHC contact or allostery. Assuming further that substitution penalties are only incurred at sites important to binding and that $P_\mathrm{S}$ and $P_\mathrm{A}$ are statistically independent, we can express co-selection factors as:
\begin{equation}
\label{eq:edAS}
    Q_\mathrm{AS}(\sigma, \sigma') \; \propto \prod_i \left[ 1 - P_\mathrm{S}(k_i, \ell) \left( 1 - P_\mathrm{A}(\Delta_i) \right) \right].
\end{equation}
 
 We set $Q_\mathrm{AS}(\sigma, \sigma') \propto e^{-d_\mathrm{AS}(\sigma, \sigma')}$ and parameterize the probabilities using substitution weights similar to those used in the previous single-factor optimization: $P_\mathrm{S}(k_i,\ell) = 1-e^{-\mathcal{S}(k_i,\ell)}$, and $P_\mathrm{A}(\Delta_i) = e^{-\mathcal{A}(\Delta_i)}$ Using these definitions, it follows that Eq.~\ref{eq:edAS} is equivalent to the family of distance metrics defined in Eq.~\ref{eq:dAS_metric}.

We found that by simultaneously accounting for both the position and identity of substitutions, $d_\mathrm{AS}$ outperforms all other distance metrics (Fig.~2B). This improved performance extends to the external validation on data from VDJdb (Fig.~2C). Further stratification of the testing data reinforces that the learned rules generalize broadly. They remain predictive even when tested on peptides differing by $\geq 6$ amino acids from the nearest training example, as well as when filtering TCRs by sequence similarity and when stratifying by HLA (Fig.~S5). Generalization to such highly dissimilar peptides has remained out of reach of previous sequence-based machine learning approaches \cite{nielsenlessons2024, meynard2024tulip}, providing empirical support for the theoretical intuition developed in Section I regarding the greater universality of co-specificity rules.

\section{Biophysical interpretation of inferred parameters}

What physical properties govern TCR specificity? Having established that the position and identity of substitutions determine the likelihood of TCR co-specificity, we turned our attention to the biopysical interpretation of the optimized parameters of the jointly optimized  $d_\mathrm{AS}$ (Eq. \ref{eq:dAS_metric}).

To judge how reliably parameters could be inferred from the available training data, we calculated their coefficients of variation across resampled training sets. We found them to be modest (Fig.~S3A,B and Fig.~S4A,B), particularly for more commonly observed substitutions (Fig.~S3C and Fig.~S4C). We expect this variability to further decrease as more data becomes available. This would potentially boost the correlation coefficients between fitted site- and substitution-weights and biophysical parameters, but some broad insights already emerge from the analysis of the available data.

\begin{figure*}
\centering
\includegraphics[scale=0.5]{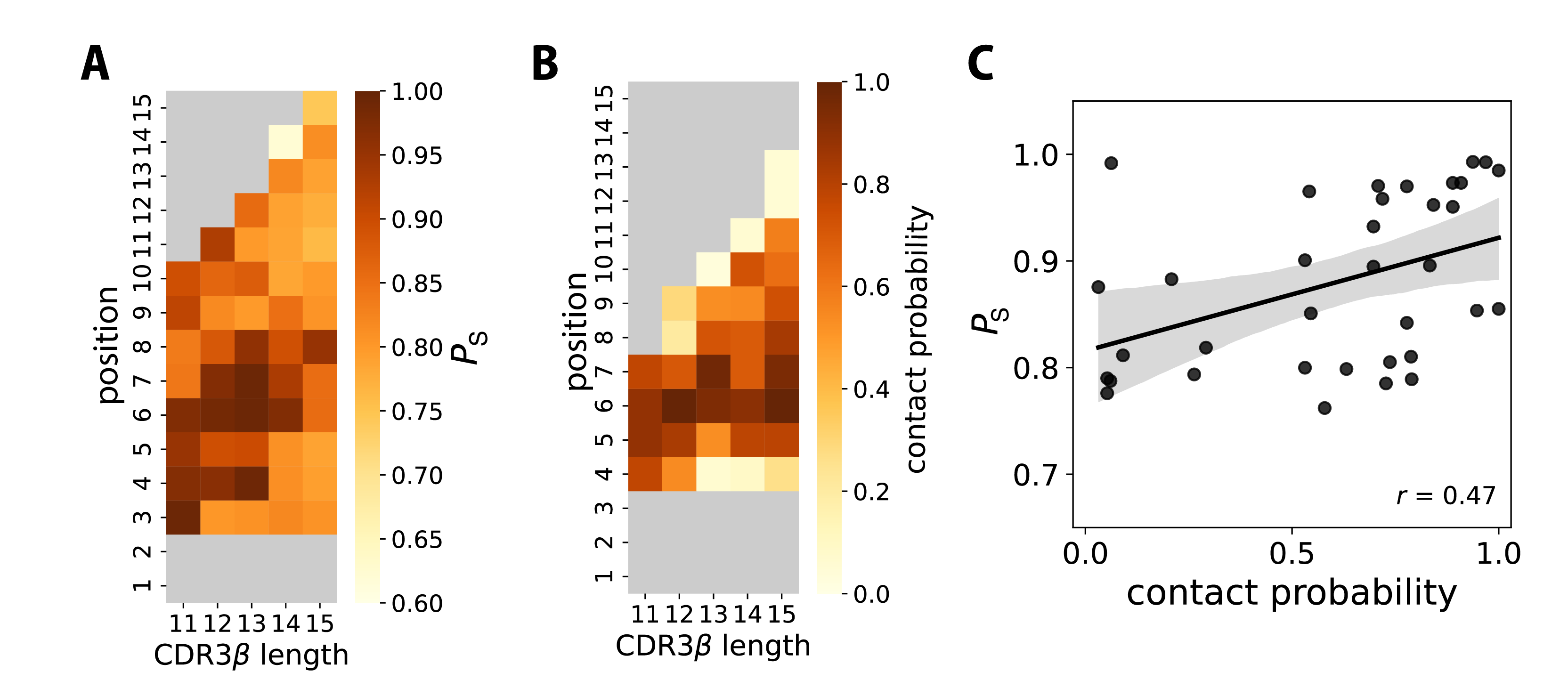}
\caption{{\bf Comparison of site-dependent weights with TCR-pMHC contact probabilities reveals the importance of non-contact sites.} \textbf{A)} Site weights, $P_\mathrm{S}$, learned from learned from joint optimization of substitution type-dependent and site-dependent weights ($d_\mathrm{AS}$). Sites that lack sufficient substitution statistics are shown in gray. \textbf{B)} Average contact probability in TCR-pMHC crystal structures (see Methods). Sites along the CDR3$\beta$ sequence that never contact the peptide in known structures are shown in gray. \textbf{C)} Scatter plot showing the correlation of both quantities across sites.} 
\end{figure*}

To test whether site weights $P_\mathrm{S}$, where higher values signify increased impact on co-specificity, are in line with their intuitive role as a proxy for contact probabilities, we calculated contact probabilities from TCR-pMHC complexes deposited in the Structural T Cell Receptor Database \cite{Leem_2017} (Fig.~3B; see Methods for details). We found that $P_\mathrm{S}$ is moderately correlated with these contact probabilities, with a Spearman's rank correlation of $r = 0.47$ (Fig.~3C). Weights $S$ learned by single-factor optimization (Fig.~S2A) were strongly correlated with jointly optimized weights $\mathcal{S}$ (Fig.~S2C), but showed a marginally weaker correlation of $r=0.45$ with contact probabilities (Fig.~S2B). In contrast to contact probabilities the optimal metric still heavily weights the edges of the CDR3 loop. This implies that substitutions of residues that do not make direct contact with the peptide substantially impact TCR specificity and should be taken into account for accurate co-specificity prediction.

\begin{figure*}
\centering
\includegraphics[scale=0.5]{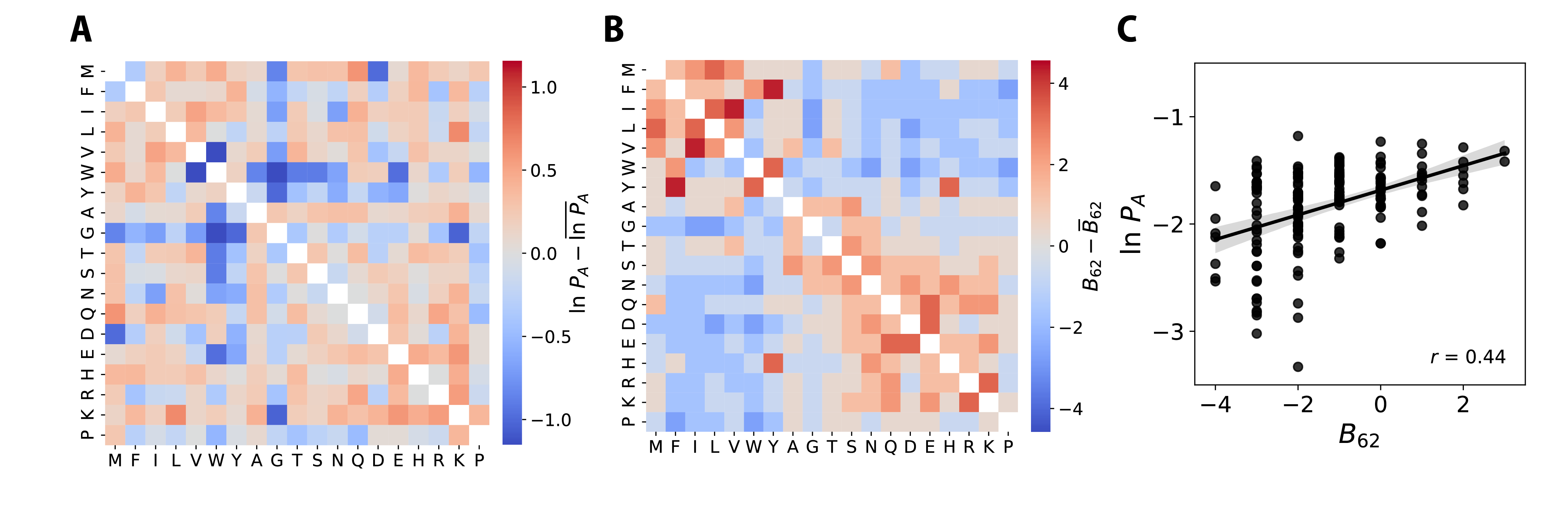}
\caption{ {\bf Optimized amino-acid substitution matrix differs from BLOSUM.} \textbf{A)} Probability of maintaining co-specificity, $P_\mathrm{A}$, learned from joint optimization of substitution type-dependent and site-dependent weights ($d_\mathrm{AS}$). Note that cysteine was removed from the optimization due to lack of substitution statistics in CDR3$\beta$ sequences. \textbf{B)} BLOSUM62 amino-acid substitution matrix. \textbf{C)} Comparison of learned $P_\mathrm{A}$ and the BLOSUM62 matrix.} 
\end{figure*}

We next compared the learned amino-acid substitution matrix $P_\mathrm{A}(\Delta)$ with prior expectations. Our results show that as expected TCR pairs differing by insertions or deletions are much less likely to be co-specific compared to those differing by amino-acid substitutions (Fig.~S1C). We further compared $\log P_\mathrm{A}$ (Fig.~4A) and the widely used BLOSUM62 matrix $B_{62}$ (Fig.~4B). Both matrices are moderately correlated with a Spearman's rank correlation coefficient of $r = 0.44$ (Fig.~4C). The jointly optimized amino-acid substitution matrix $\mathcal{A}$ was also correlated with the truncated BLOSUM62 variant used by TCRdist (Fig.~S1E), and the matrix $A$ obtained by single-factor optimization (Fig.~S1F). Single-factor optimization led to marginally weaker correlations with BLOSUM-based scores (Fig.~S1A). 

\begin{figure*}
\centering
\includegraphics[scale=0.5]{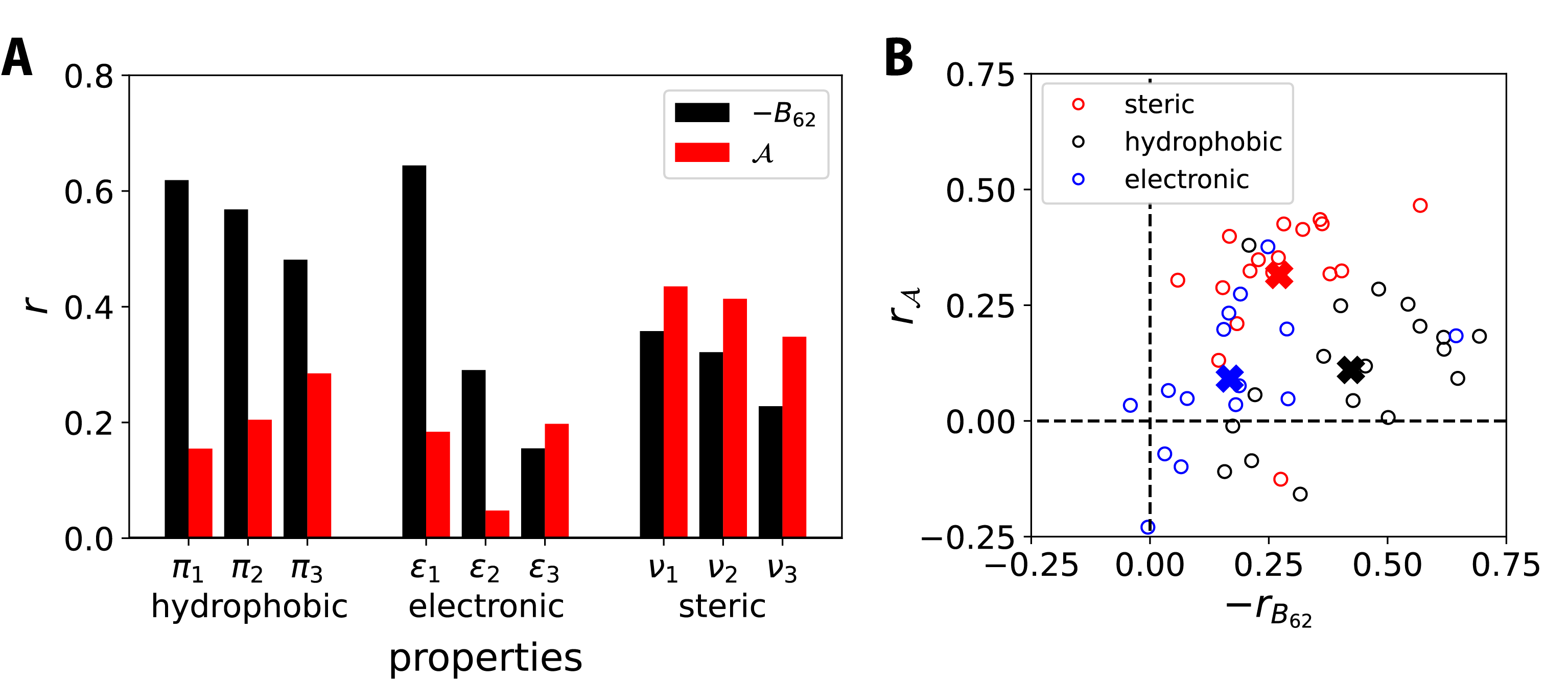}
\caption{{\bf Physical correlates of the optimized amino-acid substitution matrix highlight the contribution of shape to TCR specificity.} \textbf{A)} Spearman's rank correlation coefficient ($r$) of substitution matrices with matrices constructed from pairwise absolute difference in physical properties of amino acids. Selected hydrophobic, electronic, and steric properties were compared. Hydrophobic properties: partition coefficient in octanol/water ($\pi_1$), solvation free energy ($\pi_2$), and $dG$ of transfer from organic solvent to water ($\pi_3$). Electronic properties: polarity ($\epsilon_1$), pK of COOH on C$_\alpha$ ($\epsilon_2$), and pK of NH$_2$ on C$_\alpha$ ($\epsilon_3$). Steric properties: average volume of buried residue ($\nu_1$), normalized van der Waals volume ($\nu_2$), and STERIMOL maximum width of the side chain ($\nu_3$). \textbf{B)} Comparison of the correlation coefficients of both substitution matrices with those obtained from a panel of 50 amino-acid properties (for the complete list, see SI). Centroids for each property class are shown as "x".} 
\end{figure*}

What factors drive observed differences between $\mathcal{A}$ and $B_\mathrm{62}$? To answer this question, we identified the physical properties of amino acids that predict most closely the substitution matrices. For a given property $h(n)$ of amino acid $n$, we quantified the contribution of $h$ to the substitution matrices by calculating the matrix of absolute differences in $h$, defined as $\mathcal{H}(n,m) = |h(n) - h(m)|$. We then calculated Spearman's rank correlation coefficient between the entries of $\mathcal{H}$ and the entries of $\mathcal{A}$ and $B_{62}$ for 9 amino-acid properties.

As shown in Fig.~5A, we find that $B_{62}$ is strongly dependent on a number of hydrophobic and electronic properties of amino acids. In contrast, these properties are less correlated with the learned substitution matrix $\mathcal{A}$. Instead, $\mathcal{A}$ correlates most with the steric similarity of substituted amino acids. For instance, many of the most strongly penalized substitutions involve replacing glycine, the amino acid with the smallest and most conformationally flexible side chain. These results are compatible with our prior findings using amino-acid properties to determine optimal reduced alphabets \cite{henderson_limits_2024}. The importance of steric properties in determining the specificity of TCRs was further confirmed by analyzing a larger panel of 50 amino-acid properties (Fig.~5B) taken from Ref. \cite{Mei2005} (listed in Table S1).  

To test whether the observed decrease in the significance of hydrophobic properties—and the corresponding increase in the significance of steric properties—extends to other classes of protein–protein interactions, we analyzed thermodynamic measurements from the SKEMPI database \cite{Jankauskait2018}. Specifically, we examined which physical properties of residues (Table S1) most strongly correlate with the change in binding free energy $\Delta\Delta G$ upon mutation. As shown in Fig.~S6 steric properties are most predictive for both TCR–pMHC and antibody–antigen interactions. In contrast, for protease–inhibitor interactions, hydrophobicity emerged as the most significant factor. These results are compatible with the hypothesis that steric effects may be particularly important in interactions involving unstructured or flexible regions—such as the CDR loops in TCRs and antibodies—whereas hydrophobicity may dominate in interactions mediated by well-structured binding pockets, such as those found in proteases.

\begin{figure*}
\centering
\includegraphics[scale=0.5]{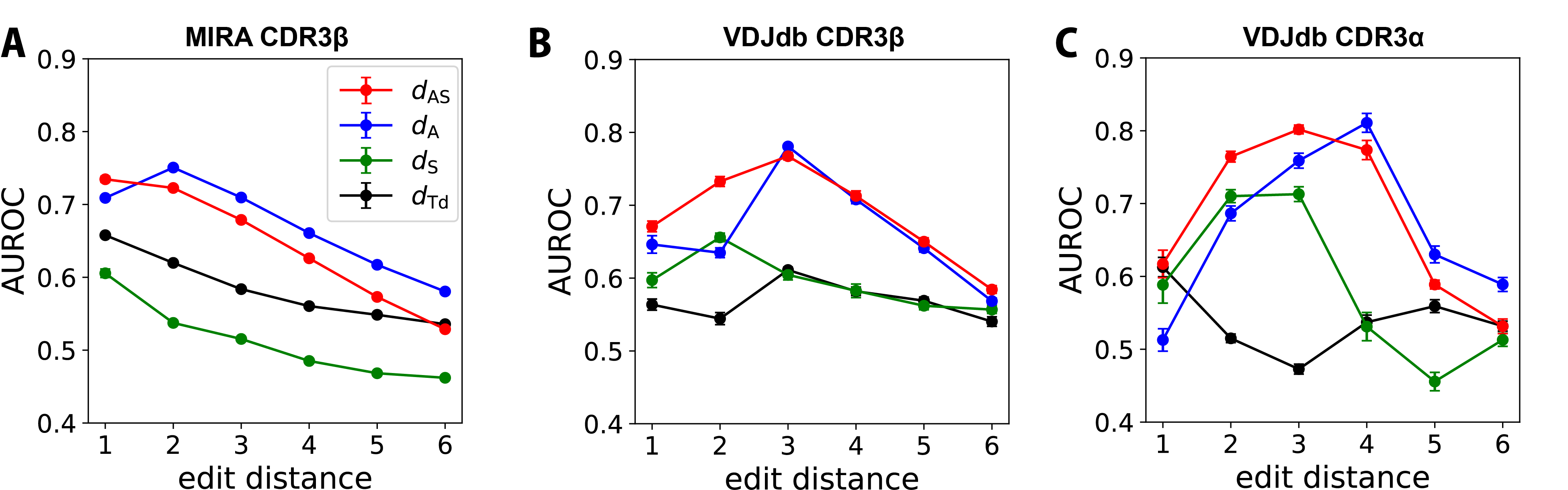}  
\caption{{\bf Learned metrics generalize to pairs of TCRs with a range of sequence similarities.} AUROC of baseline method TCRdist ($d_\mathrm{Td}$, black) and inferred site-specific metric ($d_\mathrm{S}$, green), amino-acid substitution metric ($d_\mathrm{A}$, blue), and jointly learned metric ($d_\mathrm{AS}$, red) for edit distance 1-6 TCR pairs from different datasets. \textbf{A)} MIRA test set. \textbf{B)} VDJdb CDR3$\beta$ validation set. \textbf{C)} VDJdb CDR3$\alpha$ validation set. Error bars show the standard error over resampled test batches.} 
\end{figure*}

\section{Extrapolation of learned rules}

Having learned biophysically interpretable local rules based on pairs of single-edit TCR$\beta$ sequences, we asked whether these rules can be extrapolated to more dissimilar TCR pairs and the TCR$\alpha$ chain.
To quantify this, we tested ability of trained metrics to identify co-specific sequence pairs among TCRs with an increasing number of edits on the $\beta$ chain (Fig.~6A \& B) and the $\alpha$ chain (Fig.~6C). 
This approach allowed us to evaluate the robustness and applicability of the metric across a broader range of sequence variations, providing insights into whether the local substitution rules remain effective when sequence dissimilarity increases.

We find that $d_{\mathrm{AS}}$ and $d_{\mathrm{S}}$ consistently outperform $d_{\mathrm{Td}}$ across all distances (Fig.~6A,B). These metrics optimized to take into account amino-acid substitution type ($d_{\mathrm{A}}$ and $d_{\mathrm{AS}}$) retain strong predictive power for TCR pairs with edit distances up to 3-4 edit steps. When applied to VDJdb, classification performance for certain metrics even increases for sequences that differ by two or three edits (Fig. 6B), as might be expected within the range of applicability of an independent substitution model. This relatively high predictability at moderate edit distances may, in part, reflect compound effects of multiple substitutions that make it easier to distinguish true positives from negatives. Additionally, it is important to interpret this performance in light of the labeling strategy: because we label TCR pairs from different specificity groups as negative examples, there is a greater chance of mislabeling for more similar TCRs—i.e., pairs that differ by only one or two substitutions may actually be cross-reactive, artificially decreasing the performance. At larger edit distances mislabeling of negative examples is no longer significant, as co-specific sequences become very rare. The decline in performance at even larger edit distances is likely due to the growing influence of higher-order features—such as multiple structurally distinct binding solutions \cite{mayer_measures_2023}—that are not captured by a simple independent substitution model. These findings demonstrate the strength of basic distance metrics for refining notions of similarity for TCRs, while also highlighting their limitations when comparing receptors that are highly dissimilar.

We furthermore tested the generalization of the metrics to the $\alpha$ chain using data from VDJdb. We find that the learned amino-acid substitution matrix improves upon TCRdist scoring across a range of edit distances (Fig.~6C). These results indicate that some of the local co-specificity rules are common to both $\alpha$ and $\beta$ chains. 

\section{Discussion}

Despite rapid growth in TCR-pMHC data, only a tiny fraction of the vast diversity of pairings have been measured. This sparsity complicates the discovery of generalizable rules. In this study, we addressed this problem by developing a pseudolikelihood maximization framework for optimizing notions of TCR sequence similarity. We demonstrate that it is possible to use this framework to learn rules that robustly generalize to pMHCs not used during training. Our findings highlight that, while TCR-pMHC interactions are globally complex, local sequence-based rules effectively predict co-specificity, offering a practical advance for deorphanizing TCRs of unknown specificity.

Our results challenge the use of the BLOSUM62 amino-acid substitution matrix for TCR sequence analysis. These insights confirm the independent findings of Postovskaya et al. \cite{postovskaya2025tcrblosum} using alternative approaches and data, further underscoring the need for tailored distance matrices. Specifically, we observed that hydrophobicity was a less significant factor than previously assumed, while steric similarity emerged as a more meaningful—albeit moderate—predictor of substitution impact. This shift in emphasis aligns with the classical immunological "shape space" concept \cite{perelson1979theoretical}, which posits that shape complementarity is central to receptor-ligand interactions.
Interestingly, while hydrophobicity showed limited predictive value for two-point statistics in our study, hydrophobic CDR3 residues are important in promoting the development of self-reactive T cells \cite{stadinski2016hydrophobic}, highlighting a context-dependent role.
Comparison of position-specific terms in more complex models with structural data further highlighted the importance of non-peptide-contact sites within the CDR3 to binding specificity. Such dependence on context might limit the predictive performance of TCR metrics using trimmed sequences or machine learning approaches solely taking into account contacting residues \cite{karnaukhov2024structure}, and is compatible with the presumptive role of all residues in determining overall loop geometry. 

The application of our optimization framework was limited by the available sequence data. In order to make robust inferences possible, we decided to simplify the problem setting and restrict the family of metrics used for inference. We trained on CDR3$\beta$ sequences of restricted lengths and only considered sequence pairs with matching V genes. Moreover, we only considered simple metrics with a limited number of free parameters. We assumed, for example, a factorization of site-dependent and substitution-dependent terms. With increasing data availability, including more diverse data on paired-chain TCR$\alpha\beta$ sequences and their pMHC ligands, our framework can be applied to learn more sophisticated distance metrics.

Ultimately, we envisage that it will be possible to derive a machine-learning optimized mlTCRdist metric, which can act as a drop-in replacement for TCRdist.
Additionally, we hope that the biophysical insights, robust learning framework, and data curation strategy introduced in this study can also be used to overcome current bottlenecks in the supervised contrastive training of TCR representations using deep learning architectures \cite{Leary2024,nagano2025contrastive}. Better TCR representations could significantly enhance TCR clustering for biomarker and metaclonotype discovery \cite{mayer-blackwell2021tcr}. Improved understanding of the biophysical determinants of TCR specificity could also advance rational TCR design for therapeutic applications \cite{Jones2021}.

\vspace{15pt}

\section*{Methods}

\subsection*{Training procedure}
Parameters were optimized by gradient descent, using the Adam algorithm as implemented in PyTorch \cite{paszke2017automatic}. To reduce sampling bias, the summed loss over 15 batches with equal representation of different pMHCs was simultaneously minimized, i.e., $\varepsilon(d) = \sum_{i=1}^{15}\varepsilon(d|\mathcal{P}_i,\mathcal{U}_i)$. All distance metrics were trained with a learning rate of $10^{-3}$, where 3000 steps were taken to train $d_\mathrm{S}$, 5000 steps for $d_\mathrm{A}$, and 12000 steps for $d_\mathrm{AS}$.

\subsection*{Calculation of contact probabilities}

To obtain TCR-peptide contact information, we utilized 179 publicly available TCR-pMHC complexes from the Structural T Cell Receptor Database \cite{Leem_2017}. The pairwise distances between each residue in the TCR CDR3$\beta$ loop and each peptide residue were calculated as the minimum distance between non-hydrogen atoms, as previously described \cite{milighetti2021predicting}. Residues were defined as being in contact if their pairwise distance was less than or equal to $5\ \text{\r{A}}$.

\section*{Data availability}
This study is based on publically available data from the ImmuneCODE MIRA dataset \cite{Nolan2020}, VDJdb \cite{bagaev_vdjdb_2020}, STCRdb \cite{Leem_2017}, and SKEMPI \cite{Jankauskait2018}. Source code reproducing the results reported in this manuscript is available online \cite{sourcecode}.

\section*{Acknowledgements}
AGTP was supported by Natural Sciences and Engineering Research Council of Canada (NSERC). YN and MM were supported by Cancer Research UK studentships under grants BCCG1C8R and A29287, respectively. JH was supported in part by an award (to ATM) of the National Institute for Health Research Biomedical Research Centre at NIHR UCLH BRC. The views expressed are those of the authors and not necessarily those of the NHS, the NIHR or the Department of Health. NSW was supported in part by the National Science Foundation, through the Center for the Physics of Biological Function (PHY-1734030). ATM was supported in part by funding by the Royal Free Charity. This research was also supported in part by grant no. NSF PHY-2309135 and grant no. 2919.02 by the Betty Moore Foundation to the Kavli Institute for Theoretical Physics (KITP). The authors acknowledge helpful discussions with Gian Marco Visani and Antonio Matas Gil.

\bibliography{bibliography.bib}

\clearpage
\onecolumngrid
\appendix

\setcounter{figure}{0}
\renewcommand{\thefigure}{S\arabic{figure}}
\renewcommand{\theHfigure}{S\arabic{figure}}
\renewcommand{\thetable}{S\Roman{table}}
\renewcommand{\theHtable}{S\Roman{table}}

\section*{Supplementary Figures and Tables}

\begin{figure*}[ht]
\centering
\includegraphics[width=18cm]{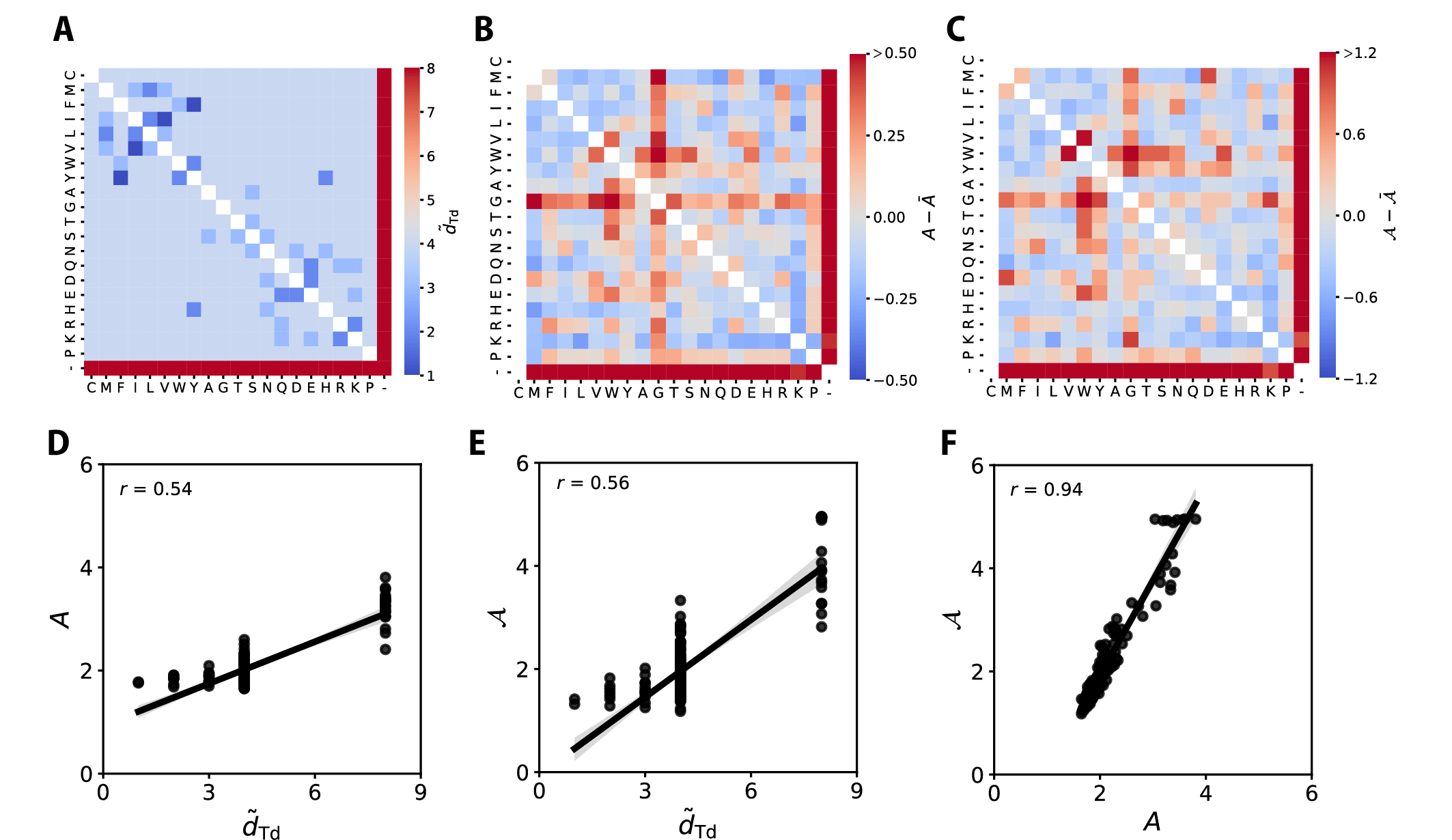}
\caption{{\bf Comparison of amino-acid substitution matrices (supplement to Fig.~4).} \textbf{A)} Substitution penalties, $\tilde{d}_\mathrm{Td}$, used in TCRdist. \textbf{B)} Centered substitution matrix $A-\bar{A}$ based on single-factor optimization ($d_\mathrm{A}$). The average value $\bar{A}$ was calculated over unlabelled sequence pairs~$\in \mathcal{U}$. \textbf{C)} Centered substitution matrix $\mathcal{A} - \bar{\mathcal{A}}$ from joint optimization ($d_\mathrm{AS}$). \textbf{D-F)} Scatter plots of substitution coefficients across the three matrices. Lines show linear fits to the data and $r$ denotes the Spearman's rank correlation coefficient.} 
\end{figure*}

\begin{figure*}
\centering
\includegraphics[width=15cm]{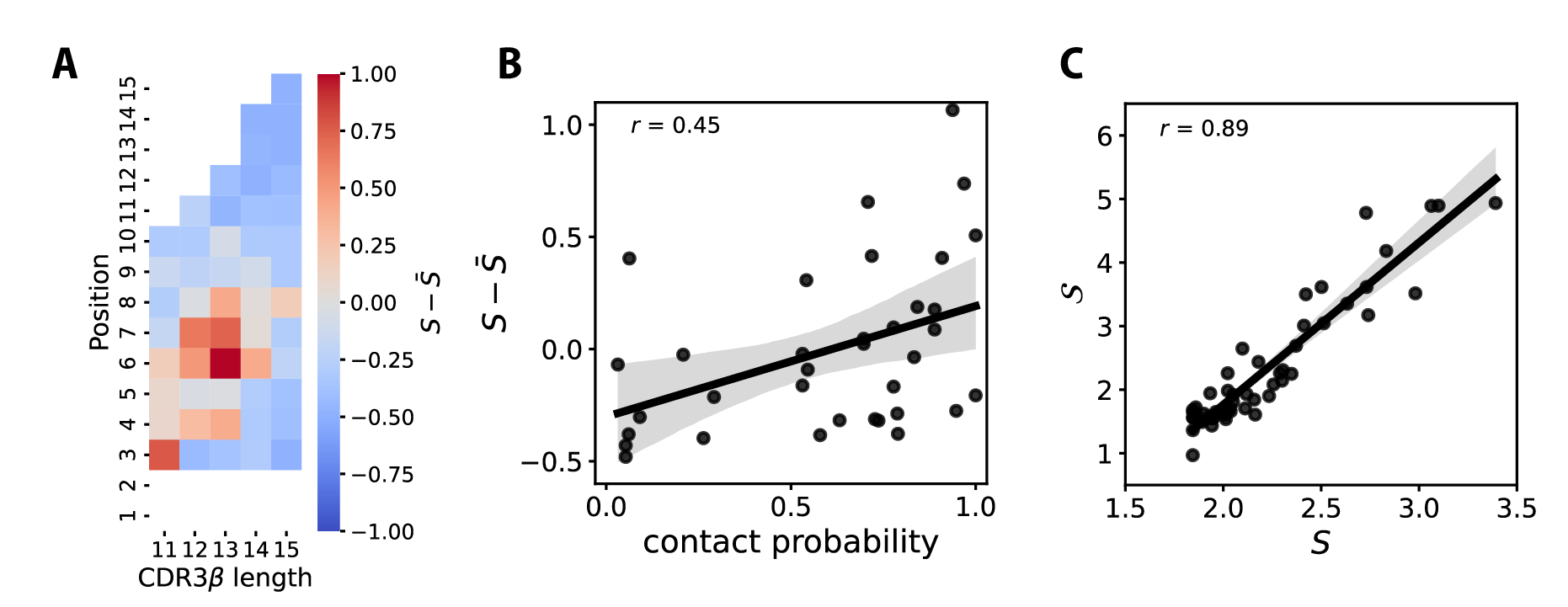}
\caption{{\bf Comparison of site weights (supplement to Fig.~3).} \textbf{A)} Centered position-dependent substitution penalties from single-factor optimization ($d_\mathrm{S}$). \textbf{B)} Comparison of $S-\bar{S}$ with contact probabilities calculated from solved structures (see Methods). \textbf{C)} Comparison of position-dependent substitution penalties $\mathcal{S}$ (associated with $d_\mathrm{AS}$) and $S$ (associated with $d_\mathrm{S}$). In B \& C, lines show linear fits to the data and $r$ denotes the Spearman's rank correlation coefficient.} 
\end{figure*}

\begin{figure*}
\centering
\includegraphics{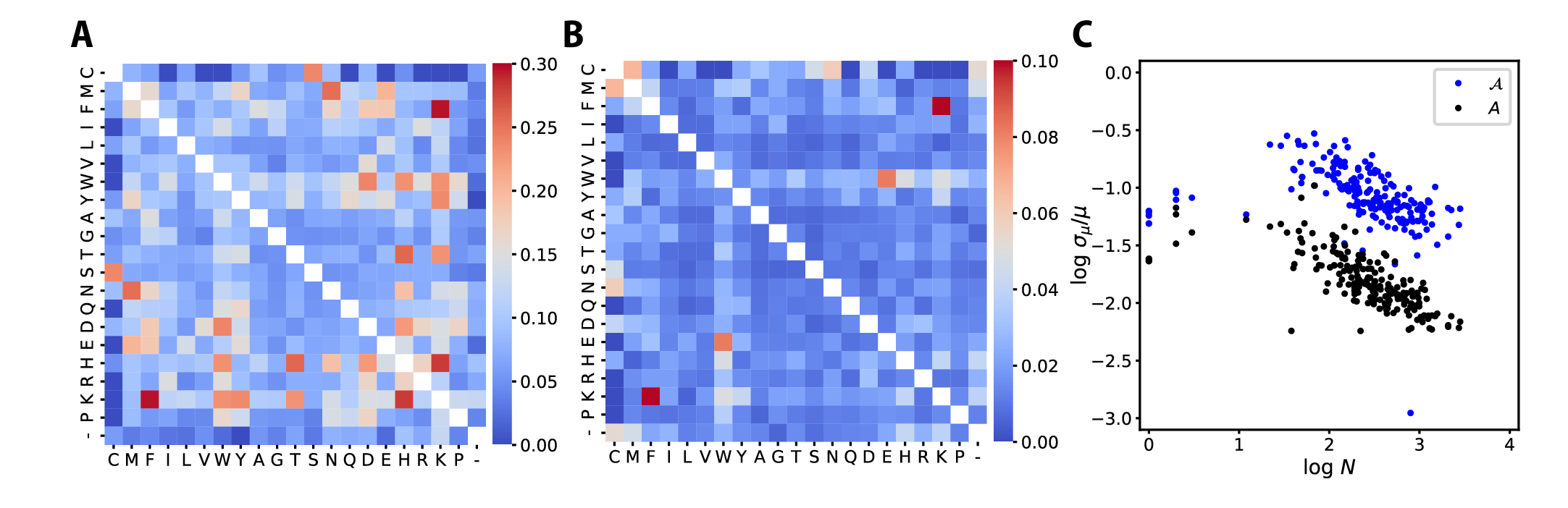}
\caption{{\bf Coefficient of variation of optimized amino-acid substitution matrices across training batches.} Matrix of the coefficients of variation, defined as the ratio of the standard error over the mean, are shown for \textbf{A)} $\mathcal{A}$ and \textbf{B)} $A$. Variability was calculated by resampling training data with replacement fiveteen times.  \textbf{C)} Variability depends on the number of available sequence pairs $N$ with a particular substitution in the training set.}
\end{figure*}

\begin{figure*}
\centering
\includegraphics{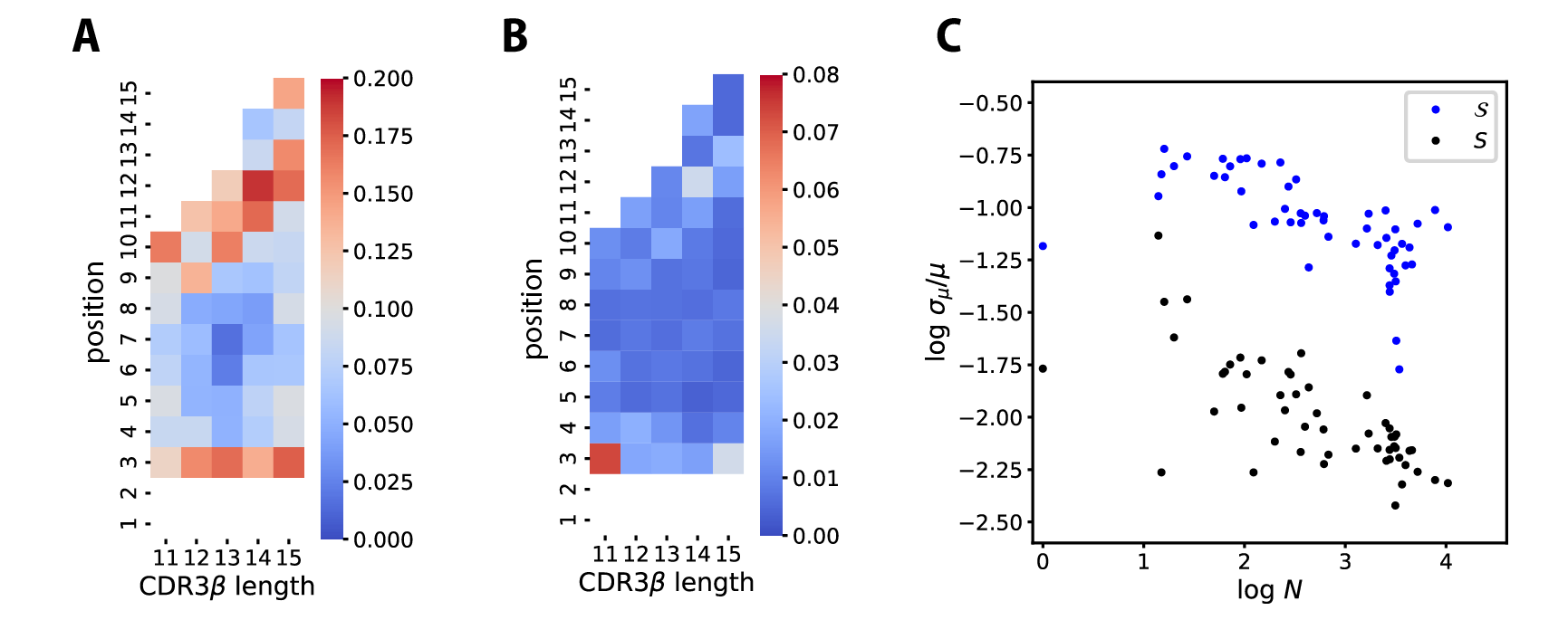}
\caption{{\bf Coefficient of variation of optimized site weights across training batches.} Matrix of the coefficients of variation, defined as the ratio of the standard error over the mean, are shown for \textbf{A)} $S$ and \textbf{B)} $\mathcal{S}$. Variability was calculated by resampling training data with replacement fiveteen times. \textbf{C)} Variability depends on the the number of available sequence pairs $N$ with a particular substitution in the training set.} 
\end{figure*}

\begin{figure*}
\centering
\includegraphics[width=12cm]{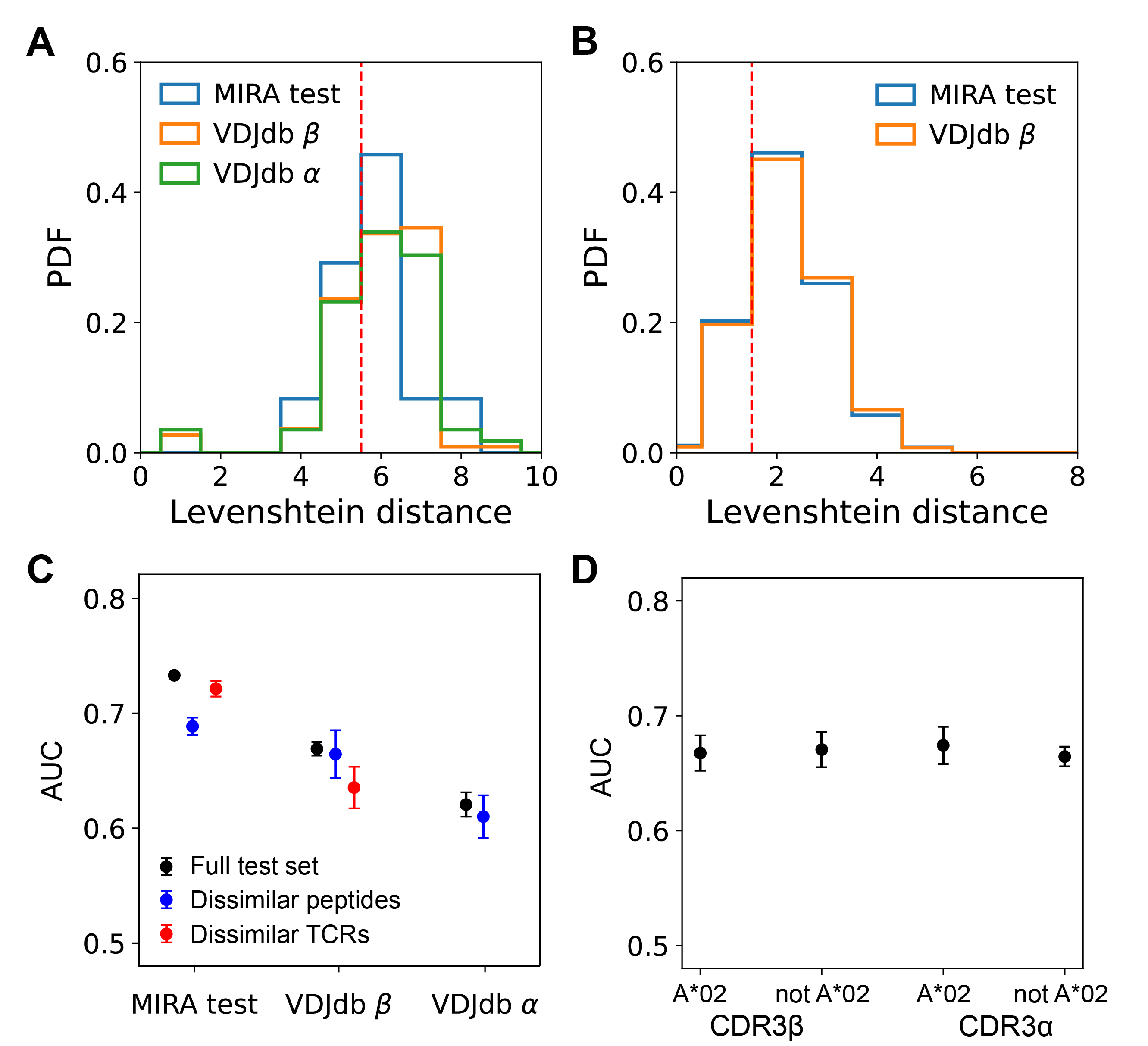}
\caption{{\bf Performance test on stratified data.} (A) Distribution of the minimum Levenshtein distances between peptide sequences in the testing set and those in the training set. Peptides with a minimum distance of at least 6 (red line) were classified as dissimilar. (B) Distribution of the minimum Levenshtein distances between CDR3$\beta$ sequences in the testing set and the training set. CDR3$\beta$s sequences with a minimum distance of at least 2 (red line) were classified as dissimilar. (C) Performance of the distance metric $d_\mathrm{AS}$ on the full testing set (black), on TCRs specific to dissimilar peptides (blue), and on dissimilar TCRs (red). (D) Performance of $d_\mathrm{AS}$ on TCR sequences from VDJdb associated with HLA-A*02 (a dominant allele in the database) and those associated with other HLA alleles. Error bars indicate the standard error across 10 test batches.} 
\end{figure*}

\begin{figure*}
\centering
\includegraphics[width=16cm]{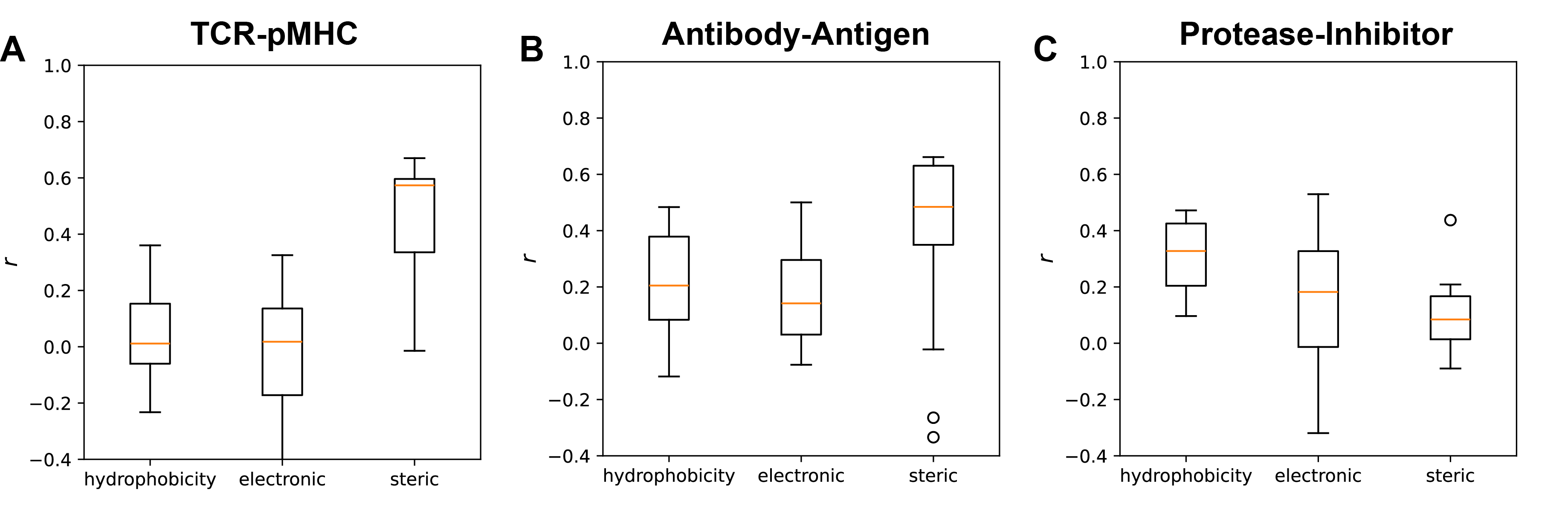}
\caption{{\bf Correlation between $|\Delta\Delta G|$ and differences in various amino acid properties.} 
This analysis uses data from SKEMPI \cite{Jankauskait2018}, a database which reports changes in thermodynamic parameters of protein-protein interactions upon mutations. The Spearman’s rank correlation coefficient was computed between $|\Delta\Delta G|$ upon mutation with the biophysical properties listed in Table SI for three classes of protein-protein interactions: (A) TCR–pMHC, (B) antibody–antigen, and (C) protease–inhibitor interactions. } 
\end{figure*}

\pagebreak
\begin{table*}[ht]
\begin{longtable}{| l | l |}
\hline
\textbf{Category} & \textbf{Description} \\
\hline
\endfirsthead
\hline
\textbf{Category} & \textbf{Description} \\
\hline
\endhead
\hline
\endfoot
\hline

\textbf{Hydrophobic Property} & 1. Retention coefficient in TFA \\
& 2. Free energy of solution in water \\
& 3. Solvation free energy \\
& 4. Melting point \\
& 5. Number of hydrogen-bond donors \\
& 6. Number of full nonbonding orbitals \\
& 7. Partition energy \\
& 8. Hydration number \\
& 9. Retention coefficient in high performance liquid chromatography (HPLC), pH 7.4 \\
& 10. Retention coefficient in HPLC, pH 2.1 \\
& 11. Partition coefficient in thin-layer chromatography \\
& 12. Retention coefficient at pH 2 \\
& 13. $R_f$ for 1-N-(4-nitrobenzofurazono)-amino acids in ethyl acetate/pyridine/water \\
& 14. $\Delta G$ of transfer from organic solvent to water \\
& 15. Hydration potential or free energy of transfer from vapor phase to water \\
& 16. $R_f$ salt chromatography \\
& 17. $\log D$, partition coefficient at pH 7.1 for acetylamide derivatives of amino acids in octanol/water \\
& 18. $\Delta G=RT\log f$, $f=$ fraction buried/accessible amino acids in 22 proteins\\
\hline
\textbf{Steric Property} & 19. Average volume of buried residue \\
& 20. Residue accessible surface area in tripeptide \\
& 21. Graph shape index \\
& 22. Normalized van der Waals volume \\
& 23. STERMIMOL length of the side chain \\
& 24. STERMIMOL minimum width of the side chain \\
& 25. STERMIMOL maximum width of the side chain \\
& 26. Average accessible surface area \\
& 27. Distance between C$_\alpha$ and centroid of side chain \\
& 28. Side-chain angle $\theta$ \\
& 29. Side-chain torsion angle $\phi$ \\
& 30. Radius of gyration of side chain \\
& 31. Van der Waals parameter $R_0$ \\
& 32. Van der Waals parameter $\epsilon$ \\
& 33. Refractivity \\
& 34. Value of $\theta$ (i) \\
& 35. Substituent van der Waals volume \\
\hline
\textbf{Electronic Property} & 36. $\alpha$CH chemical shifts \\
& 37. $\alpha$NH chemical shifts \\
& 38. A parameter of charge transfer capability \\
& 39. A parameter of charge transfer donor capability \\
& 40. Nuclear magnetic resonance (NMR) chemical shift of $\alpha$ carbon \\
& 41. Localized electrical effect \\
& 42. Positive charge \\
& 43. Negative charge \\
& 44. Polarity\\
& 45. Net charge \\
& 46. Amphipathicity index \\
& 47. Isoelectric point \\
& 48. Electron-ion interaction potential values \\
& 49. pKNH$_2$(NH$_2$ on C$_\alpha$) \\
& 50. pKCOOH(COOH on C$_\alpha$) \\
\hline
\end{longtable}
\setcounter{table}{0}
\caption{Table of amino-acid properties. This table lists the physical and chemical properties of amino acids used for comparison with substitution matrices in Fig.~5. The values were obtained from Mei et al., \textit{Biopolymers} (2005).}
\end{table*}

\end{document}